\documentclass[prd,aps,showpacs,twocolumn,preprintnumbers,nofootinbib,amsmath,amssymb,superscriptaddress, longbibliography]{revtex4-1}

\usepackage[colorlinks=true,linkcolor=blue,citecolor=blue,urlcolor=blue]{hyperref}

\usepackage{systeme}
\usepackage{mathtools}
\usepackage[inline]{enumitem}
\usepackage{graphicx}
\usepackage{upgreek}
\usepackage{combelow}

\usepackage[T1]{fontenc}
\usepackage[english]{babel}
\usepackage{xcolor}
\usepackage{cancel}
\usepackage{braket}
\usepackage{nicefrac}
\usepackage{bm}
\usepackage{bbm}
\usepackage{braket}
\usepackage{bbold}    
\usepackage{slashed}
\usepackage[normalem]{ulem}
\usepackage{array}
\newcolumntype{C}{>{$}c<{$}}

\renewcommand{\dag}{\dagger}

\newcommand{\beqn}{\begin{eqnarray}}
\newcommand{\eeqn}{\end{eqnarray}}
\newcommand{\beqs}{\begin{subequations}}
\newcommand{\eeqs}{\end{subequations}\\[-2mm]\noindent}

\newcommand{\ave}[1]{\langle #1 \rangle}

\allowdisplaybreaks

\definecolor{brickred}{rgb}{0.8, 0.25, 0.33}
\definecolor{macouleur}{RGB}{105,150,150}

\usepackage{color}
\definecolor{purple}{rgb}{0.8,0,0.6}

\begin{document}

\title{
One-quark state near a boundary of the confinement phase of QCD}

\author{Maxim N. Chernodub}
\affiliation{Institut Denis Poisson, CNRS UMR 7013, Universit\'e de Tours, Universit\'e d'Orl\'eans, 
Parc de Grandmont, Tours 37200, France}
\affiliation{Department of Physics, West University of Timi\cb{s}oara, Boulevard Vasile P\^arvan 4, Timi\cb{s}oara 300223, Romania}

\author{Vladimir A. Goy}
\affiliation{Pacific Quantum Center, Far Eastern Federal University, Vladivostok 690922, Russia}
\affiliation{Institute of Automation and Control Processes, Far Eastern Branch, Russian Academy of Science, 5 Radio Street, Vladivostok 690041, Russia}

\author{Alexander V. Molochkov}
\affiliation{Pacific Quantum Center, Far Eastern Federal University, Vladivostok 690922, Russia}
\affiliation{Beijing Institute of Mathematical Sciences and Applications,\\
No. 544, Hefangkou Village Huaibei Town, Huairou District Beijing 101408, People's Republic of China }

\author{Alexey S. Tanashkin}
\affiliation{Beijing Institute of Mathematical Sciences and Applications,\\
No. 544, Hefangkou Village Huaibei Town, Huairou District Beijing 101408, People's Republic of China }

\begin{abstract}
We discuss a one-quark state in the confinement phase near a reflective chromometallic boundary both at finite and zero temperature. Using numerical simulations of lattice Yang-Mills theory, we show that the test quark is confined to the neutral mirror by an attractive potential of the Cornell type, suggesting the existence of a mirror-bound one-quark state, a ``quarkiton.'' Surprisingly, the tension of the string spanned between the quark and the mirror is lower than the fundamental string tension. The quarkiton state exhibits a partial confinement: while the quark is localized in the vicinity of the mirror, it can still travel freely along it. Such quarkiton states share similarity with the surface excitons in metals and semiconductors that are bound to their negatively charged images at a boundary. The quarkitons can exist at the hadronic side of the phase interfaces in QCD that arise, for example, in the thermodynamic equilibrium of vortical quark–gluon plasma.
\end{abstract}

\date{\today}

\maketitle

\section{Introduction}

A fundamental property of QCD is color confinement---the empirical observation that isolated color-charged particles, such as individual quarks or gluons, are never detected under standard low-density, low-temperature environments in Nature~\cite{Greensite2020}. Instead, all physically observable states are color-singlet combinations, manifesting themselves either as mesons (quark-antiquark pairs), baryons (three-quark states), or more exotic hadronic and gluonic (glueballs) configurations~\cite{Bali:1992ru}. The confinement property disappears only under particular conditions, such as high temperatures or high baryonic densities~\cite{Laermann:2003cv}. 

The confinement of color is a nonperturbative IR phenomenon that cannot be revealed with the help of the standard perturbation theory in QCD. Confinement is believed to arise from the IR dynamics of gluons, where the strong coupling constant becomes large, making perturbative methods inadequate. As a result, the Hilbert space of asymptotic states contains only color-neutral states. The confinement property is supported also by first-principle lattice simulations, which show that in the pure-glue non-Abelian gauge theories, the potential $V_{Q \bar Q}$ between a sufficiently separated static test quark $Q$ and antiquark $\bar Q$ rises linearly~\cite{Wilson1974}, as follows:
\begin{align}
	V_{Q \bar Q}(R) \simeq \sigma R \qquad\  {\rm [large}\ Q{-}\bar Q\ {\rm separation}\ R{\rm ]}\,.
    \label{eq_V_Q_barQ}
\end{align}
The emergence of the linear potential is associated with the formation of a confining QCD string, which squeezes the electric color flux of quarks into a linear stringlike structure with a finite energy density $\sigma$ per unit string length~\cite{Greensite2020}. 

Thus, the confinement property implies that an isolated, one-quark state is not a physical state. In the stringy picture of color confinement, this statement is very transparent. The quark state does not appear alone in the vacuum, as it always comes with a chromoelectric flux tube made of gluons. One end of the chromoelectric flux tube is attached to the source of the color field, the quark itself, while the other end goes essentially to infinity because there is no antiquark at which the flux tube can terminate. Since the string carries certain energy per unit length, the total free energy of the system, ``the quark plus the attached infinitely long string,'' also becomes infinite, thus forbidding the existence of one-quark states in the confinement phase. Therefore, in the standard infinite-volume setting without boundaries, isolated one-quark states do not exist: the associated flux tube cannot terminate anywhere, its length becomes infinite, and the total free energy associated with an isolated quark diverges. 

Let us now introduce a neutral boundary and, for definiteness, take it to be a flat plane that is impermeable to quark matter but permeable for the chromoelectric flux directed along the normal to the boundary. To be more concrete, one can consider a mirror boundary that acts as a reflective mirror both for quarks and gluons. The mirror reflects back incoming quarks so that the normal component of the momentum of a quark reverses its sign upon collision with the boundary. 

For gluons, various choices of boundary conditions are possible. In this paper, we will consider a chromoelectric mirror, which is an analog of the usual metallic mirror for a photon field. Such a mirror reflects back incoming photons (gluons, in our case) and allows for the normal component of a static electric (chromoelectric) flux to penetrate through the mirror. Below, we consider the vacuum of Yang-Mills theory, where the dynamical quarks are absent so that the chromoelectric string does not break due to quark-antiquark pair creation. The static heavy quarks are introduced with the help of the Polyakov loop operator, as we discuss later. 

It is straightforward to see that the presence of a mirror allows for the existence of one-quark states near the boundary because the mentioned infinite-energy argument is no longer applicable. The chromoelectric flux tube emanating from the quark prefers to terminate at the mirror rather than at infinity. As a result, the string length becomes finite and such a one-quark state possesses a finite energy.

\begin{figure}[!htb]
    \centering
	\includegraphics[width=0.4\textwidth]{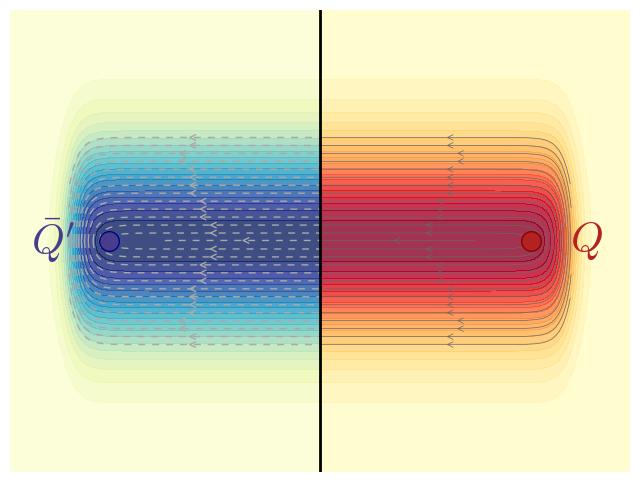} \\
    (a)\\[5mm]
	\includegraphics[width=0.4\textwidth]{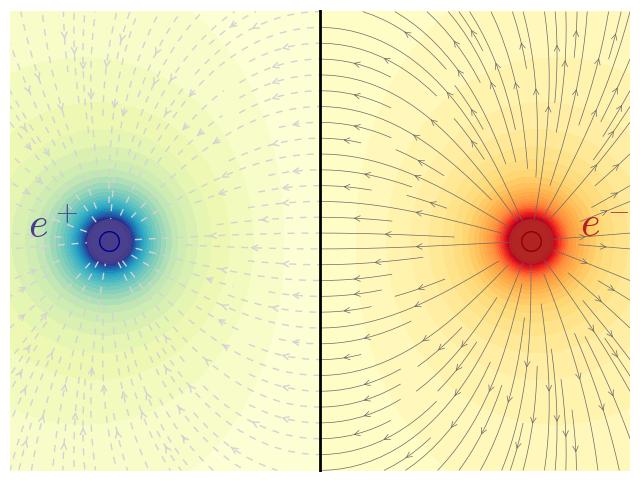} \\
    (b)
\caption{(a) Schematic visualization of the quarkiton state between a single quark $Q$ (right) and a neutral mirror (a vertical solid line). The quark is bound by a chromoelectric flux tube to the mirror by inducing an image antiquark $\bar Q$ (left). The quarkiton state (a) is a non-Abelian analog of the surface exciton (b) in which the electrons $(e^-)$ and holes $(e^+)$ form photon-mediated Coulomb-bound states with their images at the boundary.}
\label{fig_string_coloumb_compare}
\end{figure}

Thus, a minimal energy requirement for a static quark located at a fixed distance from the chromoelectric mirror boundary implies that the chromoelectric flux tube, originating at the quark ``$Q$,'' should terminate at the mirror ``$|$'' along the boundary normal. The string confines the quark to the mirror. Effectively, the heavy quark $Q$ is connected by the string to its negative image, an antiquark $\bar Q$. This configuration is visualized in Fig.~\ref{fig_string_coloumb_compare}(a)

Since the energy of the flux tube is linearly proportional to the length of the tube, this configuration would create a long-range confining linear potential between the quark and the mirror, separated by the distance $d$, as follows:
\begin{align}
	V_{Q|}(R) \simeq \sigma_{Q|} d \qquad\  {\rm [large}\ Q{-}{\rm mirror}\ {\rm separation}\ d{\rm ]}\,.
    \label{eq_V_Q_mirror}
\end{align}
Notice that in Eq.~\eqref{eq_V_Q_mirror}, the tension $\sigma_{Q|}$  of the string between the mirror and the quark is assumed to be different from the tension $\sigma$ of the string in the vacuum in the absence of boundaries~\eqref{eq_V_Q_barQ}. The investigation of the relation between $\sigma_{Q|}$ and $\sigma$ is one of the subjects of our paper. 

The long-range attraction between the quark and the mirror is supplemented, at very short distances, by the strong repulsive potential of the mirror, which is impermeable to quarks. The quark should therefore form a single-quark state bound to the mirror, the quarkiton~\cite{Chernodub:2023dok}. A single-quark quarkiton state exhibits a partial color confinement since the quark is confined to the mirror by the flux tube and it cannot move far in the normal direction. However, the quarkiton can still freely propagate along the mirror in the transverse directions.

The quarkitons, along with their gluonic counterparts, ``gluetons,'' emerge naturally as boundary (edge) states in a non-Abelian version of the Casimir effect~\cite{Chernodub:2023dok}, which describes the interaction of neutral chromoelectric or chromomagnetic mirrors in a non-Abelian vacuum~\cite{Chernodub:2018pmt, Karabali:2018ael, Chernodub:2023dok, Canfora:2024awy, Dudal:2025qqr, Karabali:2025olx}.

The quarkitons and gluetons can be regarded as non-Abelian analogs of electronic surface excitons that propagate along the boundaries of certain electronic materials such as metals, semiconductors, and organic crystals to mention a few~\cite{Agranovich2009, dean2012excitons}. The electronic surface exciton is an electrically neutral quasiparticle that emerges in semiconductors and insulators in the vicinity of the boundary of the system: an electron (or a hole) in the bulk of the material couples to its image hole (electron) state seen in the reflective boundary and forms a neutral quasiparticle confined to the boundary~\cite{mills1982surface}. 

The potential between the electron $(e^-)$ and its hole image $(e^+)$ in the mirror is given by a Coulomb interaction, shown in Fig.~\ref{fig_string_coloumb_compare}(b). The electronic surface exciton is a bound electron–hole pair whose c.m. motion and wave function are confined to the vicinity of a surface or interface.  As is clear from our description above, the quarkiton (glueton) excitation emerges essentially via the same physical mechanism: a quark (a gluon) couples to its negative image in the mirror, an antiquark (a gluon of a charge-conjugated color), and forms a neutral bound state. Another condensed-matter analog of the non-Abelian bound states is given by fractional vortices in multicomponent condensates, which are also linearly confined to the boundary~\cite{Silaev2011, Agterberg2014, Maiani2022}.

Let us stress that one of the constituents of the surface-exciton pair is a real particle (either an electron or a hole), while the other one (a hole or an electron, respectively) represents its image in the mirror. In QCD, analogously, the quarkiton is a single-quark state coupled to its antiquark image that appears in the chromoelectric mirror (and likewise for a glueton). Both gluetons and quarkitons could potentially emerge as physical objects, as they can propagate along a reflective vacuum--hadronic-interior interface. Such interfaces arise naturally within an MIT bag model~\cite{Chodos_1974je, Chodos_1974pn}. Also, stable confining-deconfining phase boundaries appear in rotating (quark-)gluon plasmas~\cite{Chernodub:2020qah, Chernodub:2022veq, Braguta:2023iyx, Braguta:2024zpi, Morales-Tejera:2025qvh}.

Certain properties of gluetons were studied in Ref.~\cite{Chernodub:2023dok}. It is instructive to compare the glueton mass with the mass of the glueball. The glueball is a color-neutral state made of gluons that exist in the bulk far from the boundaries. The mass of a ground-state glueball, by definition, defines a mass gap of the Yang-Mills system: there is no excitation with a mass lower than the ground-state glueball mass. The glueton is also a color-neutral boundary state, but it is formed by gluons bound to their image in a mirror so that the glueton is a boundary state with a restricted mobility. Unexpectedly, it appears that the mass of the glueton state is substantially lower than the mass of the glueball: the boundary glueton states exist below the mass gap set by the glueballs in the bulk~\cite{Chernodub:2023dok}. 

The ``midgap'' nature of the glueton state does not come as a surprise, given that in solid-state physics,
the states localized at the boundaries of a system (often called the edge states) can have lower masses than the bulk mass gap of the same system. This property is pertinent for the flat contacts of semiconductor structures (the Volkov-Pankratov states~\cite{volkov1985two}) as well as for the boundaries of topological insulators such as massless edge modes that appear in the spin Hall systems~\cite{hasan2010colloquium, qi2011topological}. However, contrary to the mentioned boundary states, the glueton has a nontopological origin.

The lower masses of the boundary modes can have thermodynamic implications in the physical environments where the boundaries could naturally form. A straightforward example comes from a first-phase transition that features the emergence of confining-deconfining interfaces during the nucleation processes~\cite{Kajantie:1986hq} and already mentioned phase interfaces in vortical quark-gluon plasmas~\cite{Chernodub:2020qah, Chernodub:2022veq, Braguta:2023iyx, Braguta:2024zpi, Morales-Tejera:2025qvh}. Such interfaces may also contribute via non-Abelian Casimir forces~\cite{Chernodub:2018pmt, Karabali:2018ael, Chernodub:2023dok, Canfora:2024awy, Dudal:2025qqr, Karabali:2025olx}. 

In our paper, we study the quarkiton state: a bound state of a single quark with a reflective neutral non-Abelian mirror. We aim to evaluate numerically the free energy of a single static quark in the presence of the neutral mirror and interpret the result as a potential between these objects. 

Before going into the details, it is important to notice that the computation of the quark-mirror potential is not an easy or straightforward task at zero temperature, since a single quark can only be introduced in the system with the help of the Polyakov loop $P$. This loop winds once around the imaginary time direction, which has an infinite length inversely proportional to a vanishing temperature $T \to 0$. Indeed, the free energy $F_{Q|}(T,d)$ of an isolated static color source located at a distance $d$ from the mirror is a finite quantity, so the expectation value of the Polyakov loop, $\langle P\rangle \sim e^{-F_{Q|}(T,d)/T}$, becomes exponentially suppressed~\cite{Polyakov:1978vu}. This severe suppression degrades the SNR, complicating a reliable evaluation of the expectation value $\langle P \rangle$ at low temperatures. Therefore, in our work, we perform a calculation in a range of temperatures at the confinement phase and then make an extrapolation of the quark-mirror potential $F_{Q|}(T, R)$ to the zero-temperature limit, $T \to 0$. Preliminary results, at smaller lattices and with a different renormalization convention were reported in Ref.~\cite{Tanashkin2025}. 

The structure of this paper is as follows: in Sec.~\ref{sec_YM_mirror}, we discuss Yang-Mills theory on the lattice, show how to introduce the chromoelectric mirror, and describe the parameters of the numerical simulations. Section~\ref{sec_potential} is devoted to numerical calculation of the quark-mirror potential at finite temperature and to an extrapolation of the results to the zero-temperature case. In this section, we compare the quark-mirror potential with the quark-antiquark potential in the absence of the mirror in a broad range of temperatures. The last section summarizes our conclusions.

\section{Yang-Mills theory with a mirror}
\label{sec_YM_mirror}

\subsection{Yang-Mills theory}

The action of SU(3) Yang-Mills theory in $3+1$ dimensional Minkowski spacetime has the form
\begin{equation}
    S = - \frac{1}{4} \int d^4 x \, F_{\mu\nu}^a F^{a,\mu\nu} \,,
    \label{eq_S_continuum}
\end{equation}
where $F_{\mu\nu}^a = \partial_\mu A_\nu^a -\partial_\nu A_\mu^a + g\,f^{abc}\,A_\mu^b\,A_\nu^c$ is the strength of the gluonic field $A_\mu = A_\mu^a T^a$. Here $T^a$ are the generators of the SU(3) group in the fundamental representation, normalized via ${\rm Tr}\, [T^a, T^b]=\tfrac{1}{2} \delta^{ab}$ and $f^{abc}$ are the antisymmetric structure constants of SU(3) defined as $[T^a,T^b] = i f^{abc} T^c$, with the color indices $a, b, c =1,\dots,8$. The strength of the interaction of the gluon fields is governed by the gauge coupling, $g$, which enters the field-strength tensor $F_{\mu\nu}^a$.

In our paper, we use the standard Wilson formulation of the lattice version of the Yang-Mills action~\eqref{eq_S_continuum} which is suitable for numerical simulations, as follows:
\beqn
	 S = \beta\sum_{P} \left(1 - {\mathcal P}_P \right)\,, 
     \qquad
      {\mathcal P}_P = \frac{1}{3}\mathrm{Re} \, \mathrm{Tr}\,U_P\,.
\label{eq_action}
\eeqn
The action is given by a sum over lattice plaquettes $P \equiv P_{n,\mu\nu} = \{n,\mu\nu\}$, where the indices $\mu, \nu = 1, \dots, 4$ label the directions of the axes, and the spacetime coordinate $n$ denotes a corner of the plaquette with the spacetime directions $\mu \neq \nu$ on a $4d$ Euclidean lattice. The lattice coupling $\beta = 6/g^2$ corresponds to a bare coupling of SU(3) gauge theory. 

In the continuum limit, the lattice spacing tends to zero, $a\to 0$, so that the lattice plaquette $U_{\mu\nu}(n) = U_\mu(n)U_\nu(n+\hat\mu)U^\dag_\mu(n+\hat\nu) U^\dag_\nu(n) = \exp(i a^2 F_{\mu\nu}(n) + \mathcal{O}(a^3))$ reduces to the continuum field-strength tensor~$F_{\mu\nu}$. Consequently, the lattice action~\eqref{eq_action} becomes a Euclidean version of the continuum Yang-Mills action~\eqref{eq_S_continuum}.

A static heavy quark is introduced at a spatial point~${\boldsymbol{x}}$ with the help of the Polyakov-loop operator,
\begin{align}
	P_{\boldsymbol{x}} = {\rm Tr}\, \prod_{t=0}^{N_t - 1} U_{t,{\boldsymbol{x}}}\,,
    \label{eq_P}
\end{align}
where the product is taken over a closed loop in the periodic time direction.

\subsection{Chromometallic mirror}
\label{sec_mirror}

An ideal chromometallic mirror placed at the hypersurface~${\mathcal S}$ can be modeled by the gauge-invariant non-Abelian Casimir boundary conditions imposed on gluonic fields in SU($N$) gauge theory, as follows:
\begin{equation}
	E^a_{\|}(x) {\biggl|}_{x\in {\cal S}} \! = B^a_{\perp}(x) {\biggl|}_{x\in {\cal S}} \! = 0, 
    \qquad\
    a = 1,\dots, N^2 - 1.
    \label{eq_Casimir_continuum}
\end{equation} 
These conditions imply that the tangential chromoelectric fields $E^a_i \equiv F^a_{0i}$ and normal chromomagnetic fields $B^a_i = (1/2) \varepsilon_{ijk} F^{a,jk}$ vanish at the hypersurface ${\cal S}$ of the chromometallic mirror. In our work, we consider a static mirror so that conditions~\eqref{eq_Casimir_continuum} should be applied to each time slice. These boundary conditions correspond to perfect (chromo)electric conductor conditions. The other available option is a perfect (chromo)magnetic conductor, which is considered, for example, in Ref.~\cite{Dudal:2025qqr}.

The use of the terminology ``chromometallic mirror'' is justified because the boundary conditions~\eqref{eq_Casimir_continuum}, imposed on the gluonic color fields, are identical, up to the color index $a = 1,\dots, N^2-1$, to the conditions imposed on the Abelian electromagnetic field at the surface of a perfectly conducting metal in electrodynamics. The latter corresponds to an ideal mirror for photon fields. The gauge-invariant conditions~\eqref{eq_Casimir_continuum} naturally extend this construction to non-Abelian fields, thus setting up a chromometallic mirror plate for gluons.

The particularities of the formulation of the Casimir boundary conditions on the lattice have been thoroughly discussed in our previous papers~\cite{Chernodub2016, Chernodub2022, Chernodub2023}. Below, we will briefly recall certain essential points of the construction, referring the interested reader to Ref.~\cite{Chernodub2016} for a more detailed technical description.

The Casimir boundary conditions~\eqref{eq_Casimir_continuum} in the Euclidean lattice formulation are achieved by promoting the lattice coupling in Eq.~\eqref{eq_action} to a plaquette-dependent quantity $\beta \to \beta_P$. Here, one sets $\beta_P = \lambda \beta$ if the plaquette $P$ either touches or belongs to the world hypersurface spanned by the surface ${\cal S}$ and $\beta_P = \beta$ otherwise~\cite{Chernodub2016}. The quantity $\lambda$ plays the role of a Lagrange multiplier, which, in the limit $\lambda \to \infty$, enforces the lattice version of the Casimir boundary conditions~\eqref{eq_Casimir_continuum}.

\subsection{Details of lattice simulations}

We work on asymmetric lattices, $N_t \times N_x \times N_y \times N_z$. The physical temperature is inversely proportional to the temporal extension of the lattice, $N_t$~\cite{Gattringer2010}, as follows:
\beqn
    T = \frac{1}{N_t a(\beta)}\,.
    \label{eq_temp_phys}
\eeqn
We consider different physical temperatures in the range $T \in (0.507 \dots 0.996)T_c$, where $T_c$ is the critical temperature of the deconfinement phase transition. As described below, we also make sure that the extensions of the lattice in the spatial directions are sufficiently large, $N_s \gg N_t$, with $N_s = {\rm min}(N_x, N_y, N_z)$ being the smallest spatial size of the lattice.

We study the potential between the heavy quark and the chromometallic mirror at various lattice spacings $a = a(\beta)$ to make sure that our results are insensitive to lattice artifacts associated with the lattice discretization. The physical scaling of the results is checked by increasing the temporal extent of the lattice $N_t$ and decreasing the lattice spacing $a = a(\beta)$ in such a way that the physical temperature~\eqref{eq_temp_phys} stays in the fixed range of temperatures. We also vary the physical spatial extension of the lattice to ensure reasonable independence of the results on finite volume effects. 

The critical lattice coupling $\beta_c$, corresponding to the critical temperature of the deconfinement phase transition $T_c$ is a function of the lattice extension in the temporal direction~$N_t$, cf., Eq.~\eqref{eq_temp_phys}. The critical values of the coupling $\beta_c$ are listed, for various extensions $N_t$, in Table~\ref{table_crit}. Here, we used the results of Ref.~\cite{Lucini2004} and also used our own numerical simulations to add more values that are needed for our detailed analysis.

\begin{table}[h!]
\centering
\begin{tabular}{| c | c | c | c | c |}
\hline
$N_t$ &  $\beta_c$ & $a\sqrt\sigma_0(\beta_c)$ & $a[\mathrm{fm}]$ & $T_{\rm min}/T_c$\\
\hline
5 & 5.8000 & 0.3176 & 0.163\dots 0.129 & 0.792 \\
\hline
6 & 5.8941 & 0.2612 & 0.159\dots 0.107 & 0.653 \\
\hline
7 & 5.9800 & 0.2236 & 0.151\dots 0.091 & 0.559 \\
\hline
8 & 6.0625 & 0.1947 & 0.156\dots 0.080 & 0.507 \\
\hline
9 & 6.129 & 0.1754 & 0.141\dots 0.072 & 0.507 \\
\hline
10 & 6.2 & 0.1579 & 0.127\dots0.064 & 0.507\\
\hline
\end{tabular}
\caption{Parameters of the numerical simulations. For each value of the temporal lattice extension $N_s$, we show the critical lattice coupling $\beta_c$ in the thermodynamic limit, the corresponding lattice spacing $a$ (in units of the zero-temperature string tension $\sigma_0$), and the range of the lattice spacings used (in physical units, fm). The last column shows the lowest temperature $T_{\rm min}$ (in units of the critical temperature $T_c$) used in our simulations for a particular value of $N_t$.}
\label{table_crit}
\end{table}

Most of the values of the lattice spacing, $a\sqrt\sigma_0$, for different lattice couplings $\beta$ are taken from Ref.~\cite{Athenodorou2020}. Some intermediate values are obtained using an interpolation procedure by cubic splines using the data from Ref.~\cite{Athenodorou2020}. The lattice spacings are expressed in terms of the zero-temperature fundamental string tension $\sigma_0$ at zero temperature. 

The knowledge of the dependence of the lattice spacing $a(\beta) \sqrt\sigma_0$ from Ref.~\cite{Athenodorou2020} and the values of the critical values $\beta_c = \beta(N_t)$ for each $N_t$, listed in Table~\ref{table_crit}, allow us to find the values of the coupling $\beta$ for other values of temperatures $T$ at all $N_t$. To this end, we use Eq.~\eqref{eq_temp_phys} and consider the ratio of $T_c/T$, as follows:
\begin{align}
    a\bigl(\beta(N_t)\bigr) = \frac{T_c}{T} a\bigl(\beta_c(N_t)\bigr) \,.
    \label{eq_asigT}
\end{align}
Using the cubic spline for $a(\beta)\sqrt{\sigma_0}$ as a function of the lattice coupling $\beta$, we get triplets $(T,N_t,\beta)$. We use the Monte Carlo heat bath algorithm to generate $6\times 10^5$ configurations for each point $(T,N_t,\beta)$. We remove the first $10^5$ configuration to allow for a proper thermalization of the system.

We used spatially asymmetric lattices with $N_x = N_y > N_z$. We placed a flat static chromoelectric mirror at the spatial point $z=0$. It appears that the choice of the spatially transverse extension $N_z = 30$ is enough to measure, with a satisfactory precision for our set of parameters, the effects of the mirror on the free energy of the quark. In order to achieve the thermodynamic limit, we studied the lattices with a set of large longitudinal extensions $N_x = N_y = 132, 192, 224, 254$. Together with the broad range of the lattice spacings, shown in Table~\ref{table_crit}, the variation of the spatial volume allowed us to achieve a continuum limit and ensure the independence of our results on UV (lattice spacing) and IR (lattice size) parameters of the lattice simulations.

\section{Heavy quark near a mirror}
\label{sec_potential}

\subsection{Finite temperature results and renormalization}

The unrenormalized free energy $F_{Q\bar{Q}}(T,R)$ of a static quark-antiquark pair $Q \bar Q$ separated by the distance $R = |{\boldsymbol{R}}|$ at temperature $T$ is given by the correlator of two Polyakov loops~\cite{Gattringer2010}:
\begin{align}
    \ave{P_{{\boldsymbol{x}}}P_{{\boldsymbol{x}}+{\boldsymbol{R}}}^\dagger} = e^{- F_{Q\bar{Q}}(T,R)/T}\,.
    \label{eq_FEqq}
\end{align}
The expectation value of this operator is taken in a sufficiently large system, implying that the distance between the quark and the antiquark is much smaller than the size of the system in any spatial direction. In order to improve the statistics and enhance the SNR ratio, the expectation value~\eqref{eq_FEqq} is averaged over all spatial points $\boldsymbol{x}$ and over all orientations of the interparticle distance $\boldsymbol{R}$, compatible with the lattice symmetries. 

In the confinement phase, $T < T_c$, the quark-antiquark potential is a linearly rising function of the distance between the constituents of the pair~\eqref{eq_V_Q_barQ}. Therefore, the correlator~\eqref{eq_FEqq} diminishes quickly as the size of the pair $R$ increases, $\ave{P_{{\boldsymbol{x}}}P_{{\boldsymbol{x}}+{\boldsymbol{R}}}^\dagger} \sim \exp\{- \sigma R/T\}$.  For a single quark, $\ave{P_{{\boldsymbol{x}}}} \sim \exp\{- \sigma L/T\}$, where $L$ is the spatial size of the system. Therefore, the expectation value of the Polyakov loop vanishes in the thermodynamic limit. 

In our work, we introduce a static flat chromoelectric mirror at the position $z=0$ that extends through the whole system in the directions $x$ and $y$. The plate is, effectively, infinite, as it is closed via the periodic boundary conditions of a very large system. We use the lattice version of the perfect-chromometallic boundary condition~\eqref{eq_Casimir_continuum} following the procedure given in Sec.~\ref{sec_mirror}. 

According to our earlier arguments discussed in the Introduction, a single quark can possess a finite free energy near the mirror even in the thermodynamic limit. The expectation value of the single Polyakov loop,
\begin{align}
    \ave{P_{{\boldsymbol{d}}}}_{|} = e^{- F_{Q|}(T,d)/T}\,,
    \label{eq_free_en}
\end{align}
gives us the unrenormalized free energy $F_{Q|}(T,d)$ of the single quark at the distance $d = |\boldsymbol{d}|$ near the mirror. The free energy is identified with the quark-mirror potential, $F_{Q|}(T,d) \equiv V_{Q|}(T,d)$, which is expected to provide a linear confining force between the quark and the mirror~\eqref{eq_V_Q_mirror} at large quark-mirror separations~$d$. 

\begin{figure}[ht]
\centering
  \includegraphics[width=0.975\linewidth]{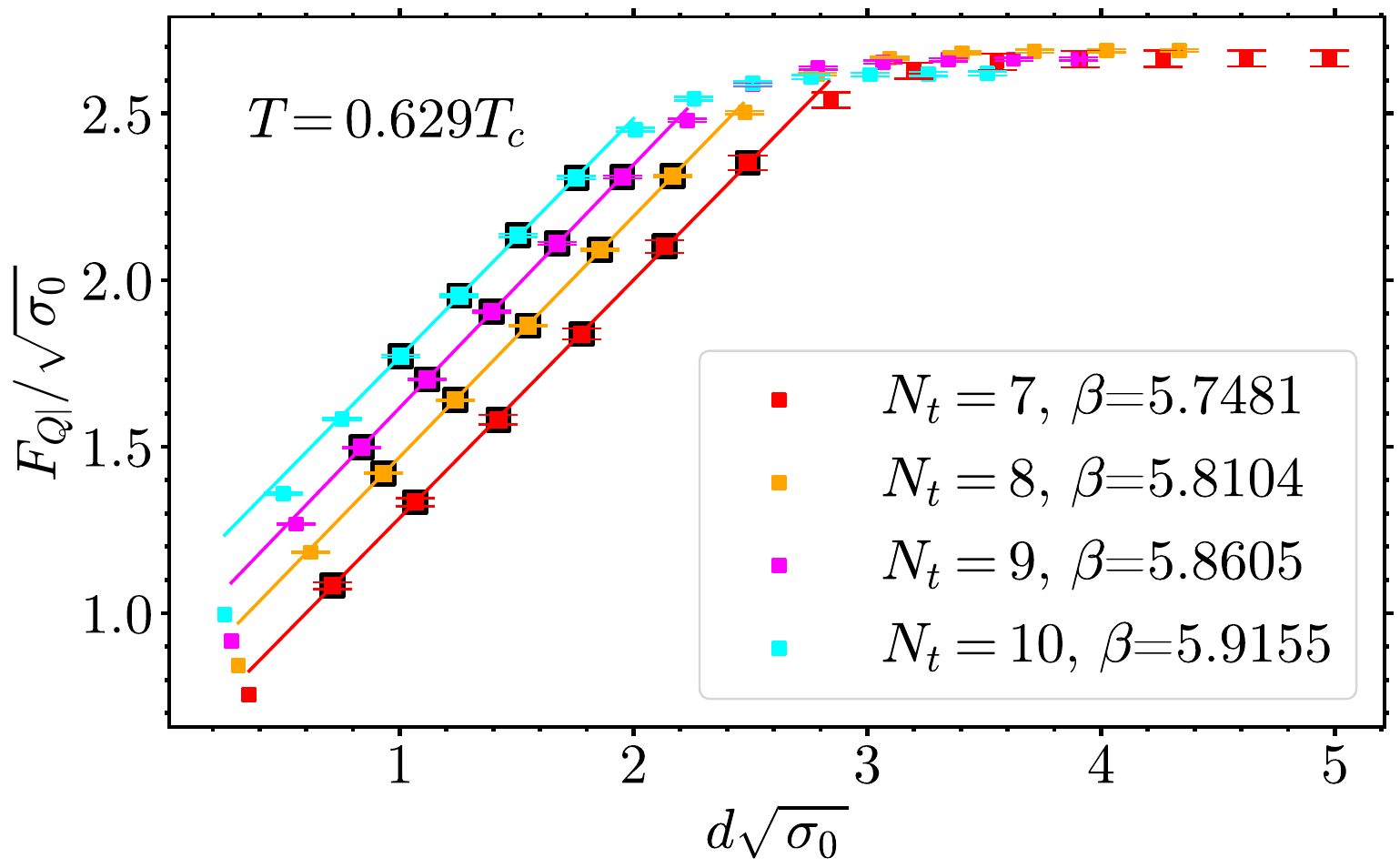}
  \caption{Unrenormalized free energy of a static heavy quark as a function of distance to the mirror $d$ at temperature $T=0.629T_c$ for several values of the temporal extension $N_t$ of the lattice, which correspond, via Eq.~\eqref{eq_temp_phys}, to different UV cutoffs (lattice spacings) $a$. The linear slopes of potentials are marked by the straight lines.
  The physical scale is given by the zero-temperature fundamental string tension $\sigma_0$. The calculation is performed at $N_t \times 254 \times 254 \times 30$ lattices.}
  \label{fig_FE_0629_phys}
\end{figure}

In Fig.~\ref{fig_FE_0629_phys}, we show an example of the single-quark free energy near the mirror, which is computed numerically using the single-point Polyakov loop~\eqref{eq_free_en} at the temperature $T=0.629T_c$. We present the data for various extensions of the lattice in the temporal direction $N_t$, which allows us to calculate the same physical quantity at different lattice spacings $a$. Our numerical data show that while the quark-mirror potential is affected by the UV cutoff $a$, the linear coefficients of the straight slopes, marked by the straight lines in Fig.~\ref{fig_FE_0629_phys}, are largely independent of the value of the lattice spacing~$a$. Also, at a sufficiently large separation $d$ from the mirror, the single-quark free energy flattens and collapses to the same value for all values of  the cutoff~$a$. The value of the asymptotic Polyakov loop depends on the spatial size of the system. The $d \to \infty$ plateau vanishes in the thermodynamic limit in the consistency of the color-confining property.

\begin{figure}[!htb]
    \centering
\begin{tabular}{c}
	        \includegraphics[width=0.45\textwidth]{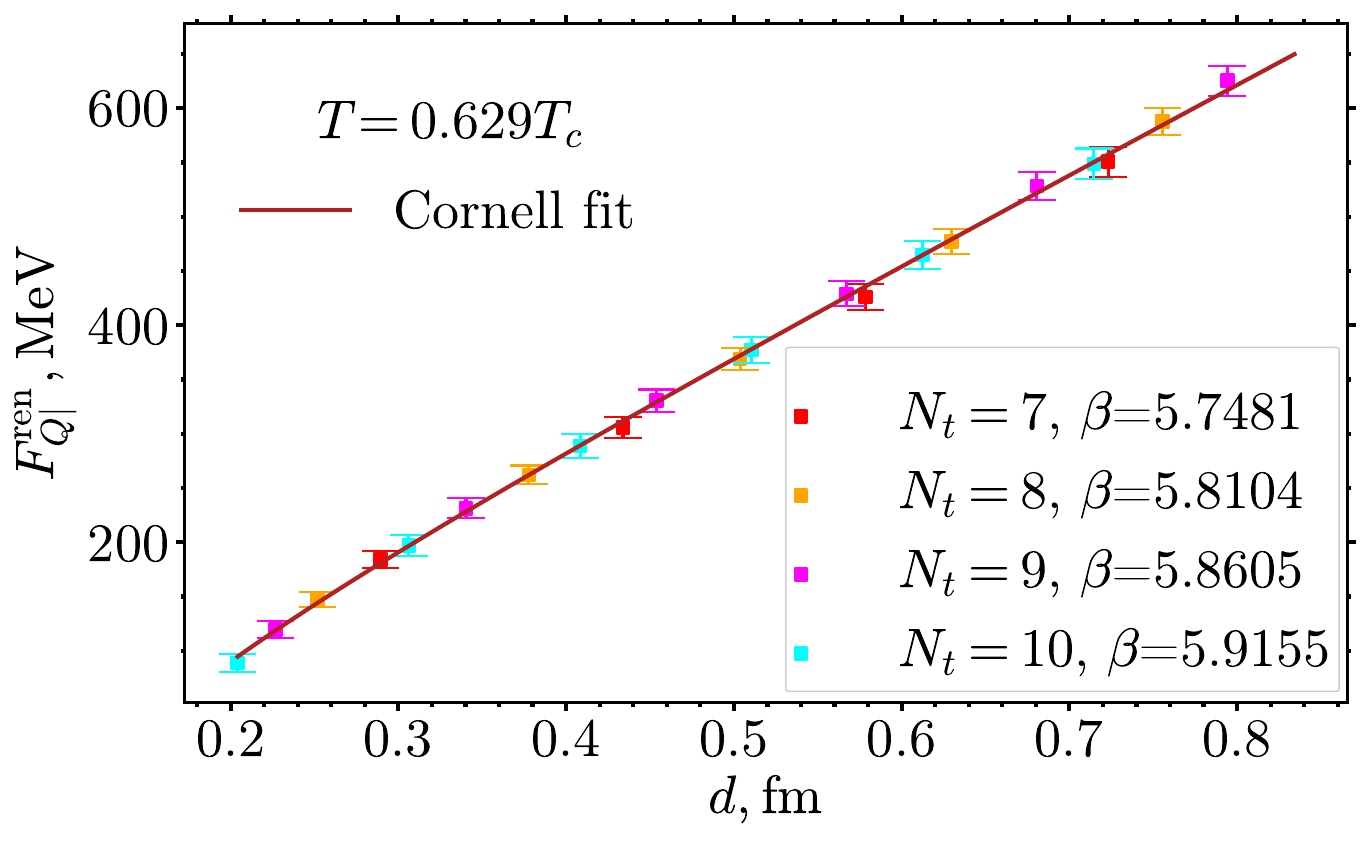}\\
            (a)\\[2mm]
	        \includegraphics[width=0.46\textwidth]{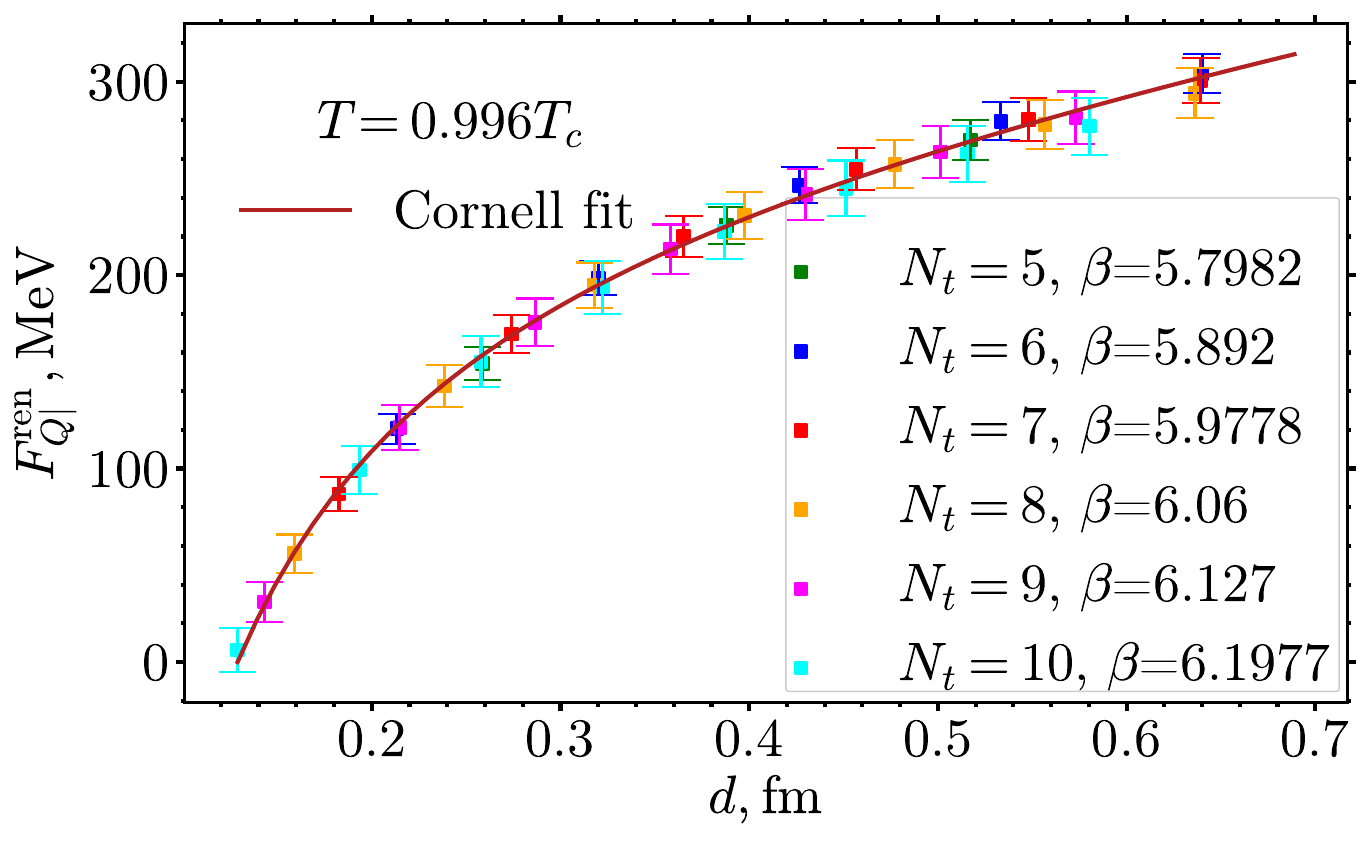}\\
	        (b)
\end{tabular}
	\caption{Renormalized free energy of a single heavy quark at the distance $d$ to the chromoelectric mirror at the temperatures (a) $T = 0.629 T_c$ and (b) $T = 0.996 T_c$ for various lattice spacings controlled by the temporal extension of the lattice $N_t$. The solid lines show the best fits by the Cornell potential~\eqref{eq_cornell_fit}. We use the $N_t \times 254 \times 254 \times 30 $ lattices.}
	\label{fig_cornell_plot}
\end{figure}

Figure~\ref{fig_FE_0629_phys} indicates that the single-quark potential is affected by the UV effects that provide an additive correction to the quark self-energy. Therefore, the free energy needs to be renormalized similarly to the renormalization of the quark-antiquark potential in the absence of the boundary~\cite{Kaczmarek:2002mc}. In our paper, we perform the renormalization of the quark-mirror interaction following the idea of matching the low-distance behavior of the potentials discussed in Ref.~\cite{Kaczmarek:2002mc}. 

The UV contribution additively renormalizes the free energy, shifting the lattice data by a finite quantity, $F_{Q|}(d,\beta) = F_{Q|}^\mathrm{ren}(d) + \Delta F(\beta)$, where $F_{Q|}^\mathrm{ren}(d)$ is the physical (renormalized) free energy of a heavy quark that does not depend on the lattice coupling $\beta$. For the case of the particular temperature $T=0.629T_c$, shown in Fig.~\ref{fig_FE_0629_phys}, the lattice free energy of the heavy quark can be renormalized by additively shifting the lattice free energy by the simple linear function $\Delta F(\beta) = a_0 + a_1 \beta$, with parameters $a_0 = -15.68(24)$ and $a_1 = 2.83(4)$. We found that the same linear renormalization applies for all considered temperatures ranging from 0.507$T_c$ to 0.996$T_c$, with the coefficients $a_0$ and $a_1$ that depend on the physical temperature $T$. To show the data in physical units, we adopted the physical values for the fundamental string tension at zero temperature $\sigma_0 = (485 \,{\rm MeV})^2$~\cite{Athenodorou2020}.

After subtracting the UV contribution from the lattice data and excluding the flattening long-distance tails, we finally get the renormalized free energy $F_{Q|}^\mathrm{ren}(d)$. 
In Fig.~\ref{fig_cornell_plot}, we show the examples of the renormalized free energies of quarks near the mirror for two representative temperatures, far from the transition point, $T = 0.629T_c$ and very close to the transition, $T = 0.996T_c$. 

The data for the renormalized free energy perfectly arrange to the single curve, thus indicating the independence of our data on the UV lattice cutoff $a = a (\beta)$. We also checked the insensitivity of the results on the spatial lattice size $L_s = N_s a$. 

We found that the renormalized free energy can be accurately fitted by the Cornell potential~\cite{Eichten:1978tg},
\beqn
    F_{Q|}^{\rm ren}(T,d) = - \frac{\alpha_{Q|}(T)}{d} + \sigma_{Q|}(T) d + F_0(T)\,,
    \label{eq_cornell_fit}
\eeqn
in a wide range of distances from the mirror. The examples of the fits are shown in Fig.~\ref{fig_cornell_plot} by solid lines. 

The first term in the Cornell potential~\eqref{eq_cornell_fit} represents the perturbative Coulomb ($\propto 1/d$) potential, which appears due to a gluon exchange between the heavy quark and its image in the mirror. This term, given by a perturbative interaction between the quark and the wall, is related to the strong coupling $\alpha_s$ at short distances. For the temperatures indicated in Fig.~\ref{fig_cornell_plot}, the best fits give $\alpha_{Q|} = 0.056(9)$ for $T=0.629T_c$ and $\alpha_{Q|} = 0.181(5)$ for $T=0.996T_c$, respectively. 

The second term in Eq.~\eqref{eq_cornell_fit} corresponds to the nonperturbative linear ($\propto d$) growth of the free energy due to the formation of the confining string between the heavy quark and the neutral mirror. This term is important at large distances. For the data shown in Fig.~\ref{fig_cornell_plot},
the best-fit values for the string tension in Eq.~\eqref{eq_cornell_fit} are $\sigma_{Q|}  = 0.688(10)\sigma_0$ for $T = 0.629 T_c$ and $\sigma_{Q|}  = 0.136(8)\sigma_0$ for $T = 0.996 T_c$. The last term in Eq.~\eqref{eq_cornell_fit} is an inessential additive factor. 

The excellent fits of the free energy of an isolated quark near a neutral mirror by the Cornell potential~
\eqref{eq_cornell_fit} indicate that the quark indeed generates an antiquark image in a mirror and gets attracted to it. At large distances from the mirror $d$, the linear attraction of the heavy test quark to the mirror implies that the quark is indeed confined to the chromoelectric mirror, thus supporting our previous conclusion on the formation of the nonperturbative quarkiton boundary state.

\subsection{Extrapolation to zero temperature}

\begin{figure}[ht]
\centering
  \includegraphics[width=0.975\linewidth]{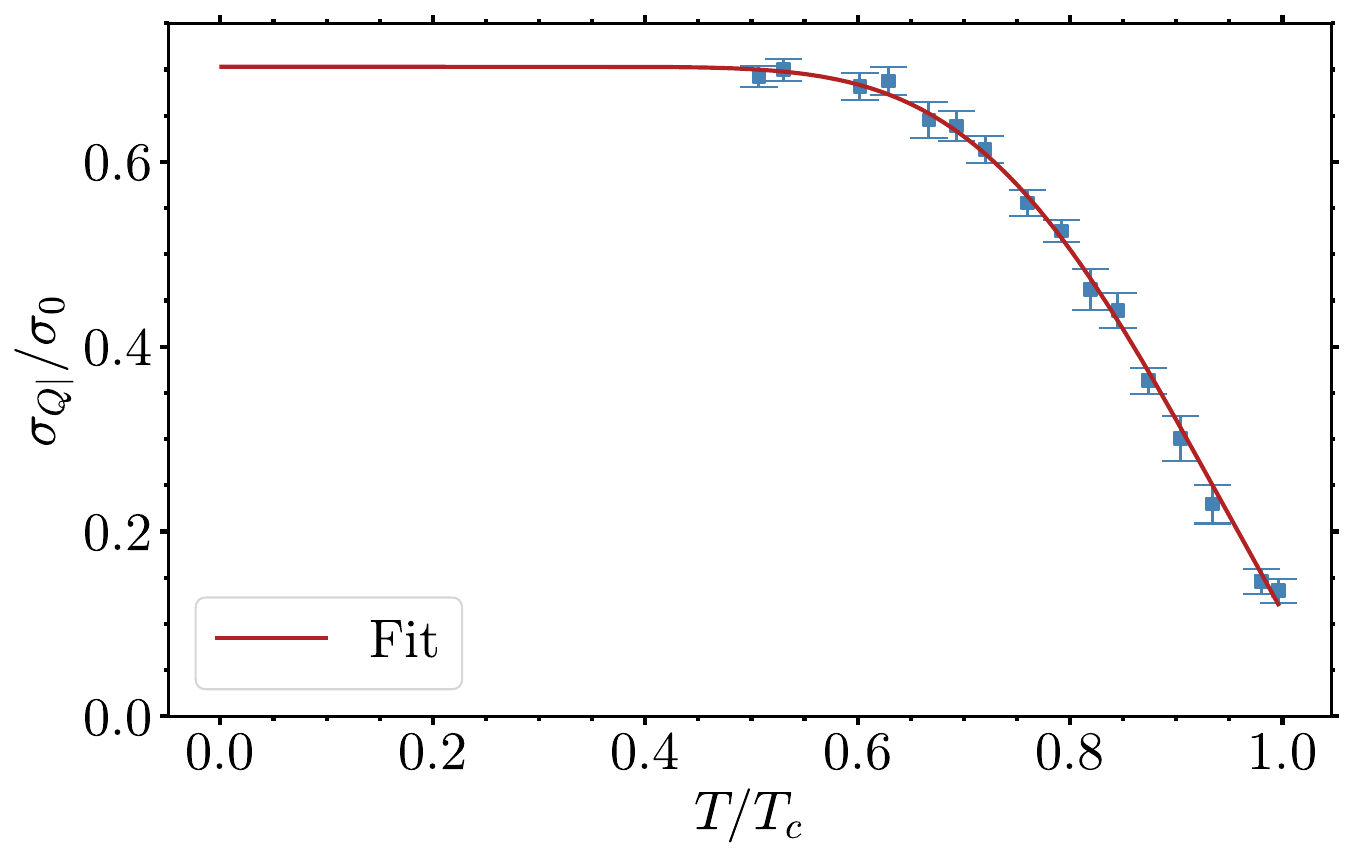}
  \caption{Quarkiton string tension $\sigma_{Q|}$, which determines the long-distance linear attraction~\eqref{eq_cornell_fit} between a single heavy quark $Q$ and a neutral chromometallic mirror, as a function of temperature $T$. The solid line represents the best fit of the data by function~\eqref{eq_sigmaQ_fit}.}
  \label{fig_sigmaQ_plot}
\end{figure}

We found that the free energy of the heavy quarks can be excellently described by the Cornell potential~\eqref{eq_cornell_fit} in the whole range of studied temperatures, $0.507 T_c < T < 0.996 T_c$. In particular, the long-distance tail of the Cornell potential~\eqref{eq_cornell_fit} corresponds to a linear  ``stringy'' potential that attracts the quark to the wall. Our numerical data show that the corresponding quarkiton string tension $\sigma_{Q|}$, presented in Fig.~\ref{fig_sigmaQ_plot}, is a diminishing function of temperature $T$.

The quarkiton string tension does not vanish as temperature approaches the critical temperature from the direction of the confinement phase, similarly to the behavior of the fundamental string tension between a quark and an antiquark~\cite{Cardoso2012}. At the low temperature limit, the string tension is approaching a plateau at a nonzero value. We found that at the whole range of temperatures, the quarkitonic string tension $\sigma_{Q|}(T)$ can be well described by the following function:
\begin{align}
    \frac{\sigma_{Q|}(T)}{\sigma_0} = c_1 \tanh \Bigl(t_0 \frac{T_c}{T} + c_0 \Bigr)\,,
    \label{eq_sigmaQ_fit}
\end{align}
where $c_0$, $c_1$ and $t_0$ are the fitting parameters. 

The best fit of our data~\eqref{eq_sigmaQ_fit} is shown in Fig.~\ref{fig_sigmaQ_plot} by the solid line. The fit has perfect $\chi^2$ statistics, $\chi^2/{\rm d.o.f.} = 1.02$, with the best-fit coefficients $c_0 = -2.806(90)$,  $c_1 = 0.703(5)$ and $t_0 = 2.969(85)$. From this best fit, we can estimate the string tension of a quarkiton at zero temperature, $T=0$ which is dictated by the $c_1$ coefficient:\footnote{In Ref.~\cite{Tanashkin2025}, we worked at smaller lattices and used another definition of the ratio. Result~\eqref{eq_R} represents an improved estimation of the quarkitonic string tension.} 
\begin{align}
		\mathcal{R}_{Q|} = \lim_{T \to 0}  \frac{\sigma_{Q|}(T)}{\sigma_{Q \bar Q}(T)} = 0.703(5)\,.
    \label{eq_R}
\end{align}
However, this method could be criticized as inexact since the low-temperature plateau, shown in Fig.~\ref{fig_sigmaQ_plot}, is not sufficiently developed in the sense that there are not enough data points at the plateau. Unfortunately, below the lowest studied temperature, the data for the expectation value of the Polyakov loop become very noisy, thus prohibiting us from studying more points at the suspected plateau. 

\begin{figure}[ht]
\centering
  \includegraphics[width=0.975\linewidth]{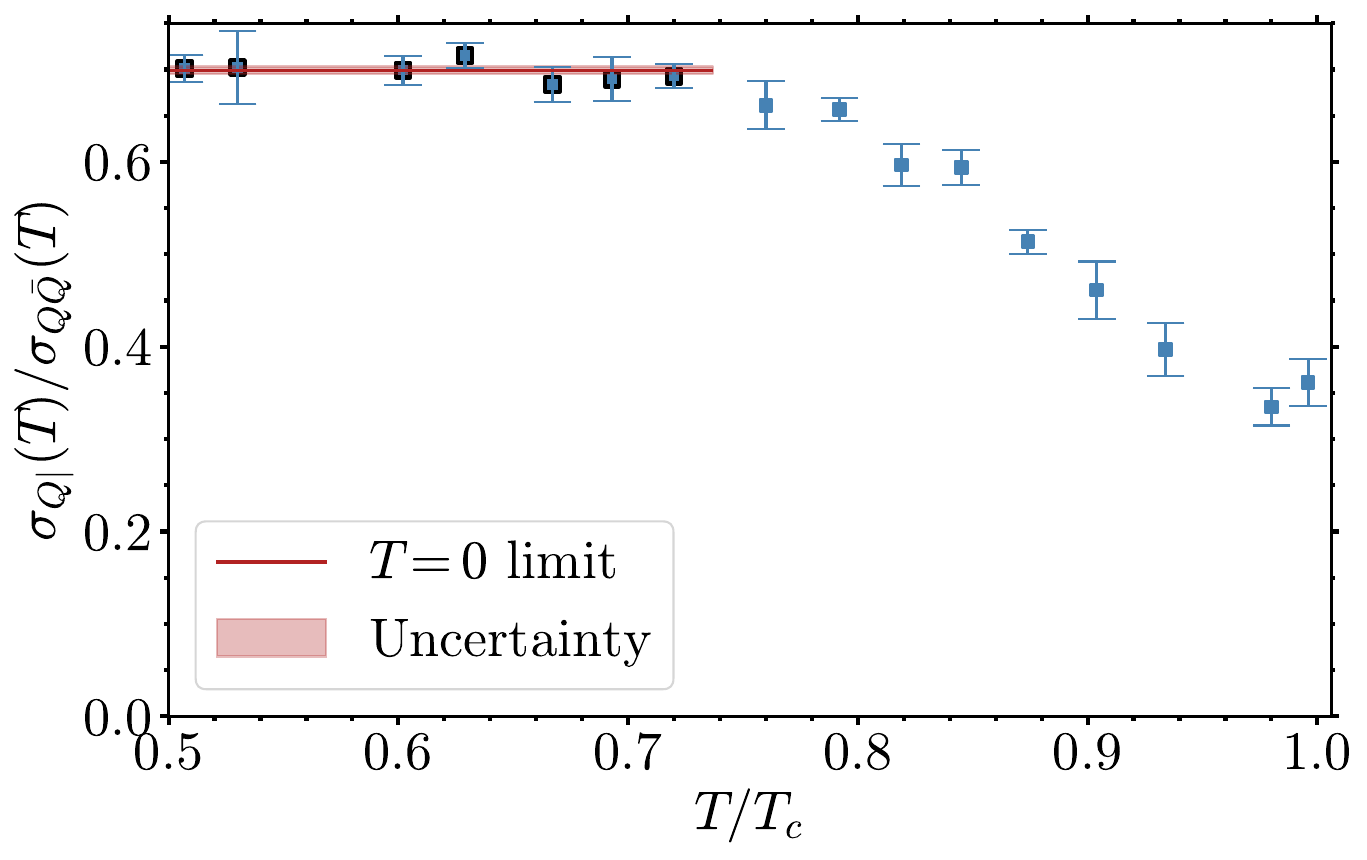}
  \caption{Ratio of the quarkitonic string tension to the fundamental string tension, $\sigma_{Q|}(T)/\sigma_{Q\bar Q}(T)$, for different temperatures. The average of the plateau in the range of temperatures $T \leqslant 0.75 T_c$ is shown by the solid line, with the uncertainty shown by the shaded region.}
  \label{fig_ratio_sigQ_sig}
\end{figure}

However, there is yet another way to estimate the low-temperature asymptotics of the quarkiton string tension. To this end, we plot in Fig.~\ref{fig_ratio_sigQ_sig} the ratio of the quarkitonic string tension and the fundamental string tension, taken from Ref.~\cite{Cardoso2012}, at the same values of the finite temperature. The plot shows that the quarkitonic string tension close to the phase transition approaches a fraction of the fundamental string tension taken at the same temperature:
\begin{align}
	\lim_{T \to T_c^-} \frac{\sigma_{Q|}(T)}{\sigma_{Q \bar Q}(T)} \simeq 0.35\,.
\end{align}
At the low-temperature limit, Fig.~\ref{fig_ratio_sigQ_sig} shows a well-developed plateau at $T \leqslant 0.75 T_c$. Averaging the data over the plateau, we arrive at the following value of the quarkitonic string tension at the low-temperature limit, $\mathcal{R}_{Q|} = \lim_{T \to 0}\sigma_{Q|}(T)/\sigma_{Q \bar Q}(T) = 0.699(4)$. This value is perfectly consistent with our earlier estimation~\eqref{eq_R}, based on the fitting of the finite-temperature behavior of the quarkionic string tension by the empirical dependence~\eqref{eq_sigmaQ_fit}.

\begin{figure}[ht]
\centering
  \includegraphics[width=0.975\linewidth]{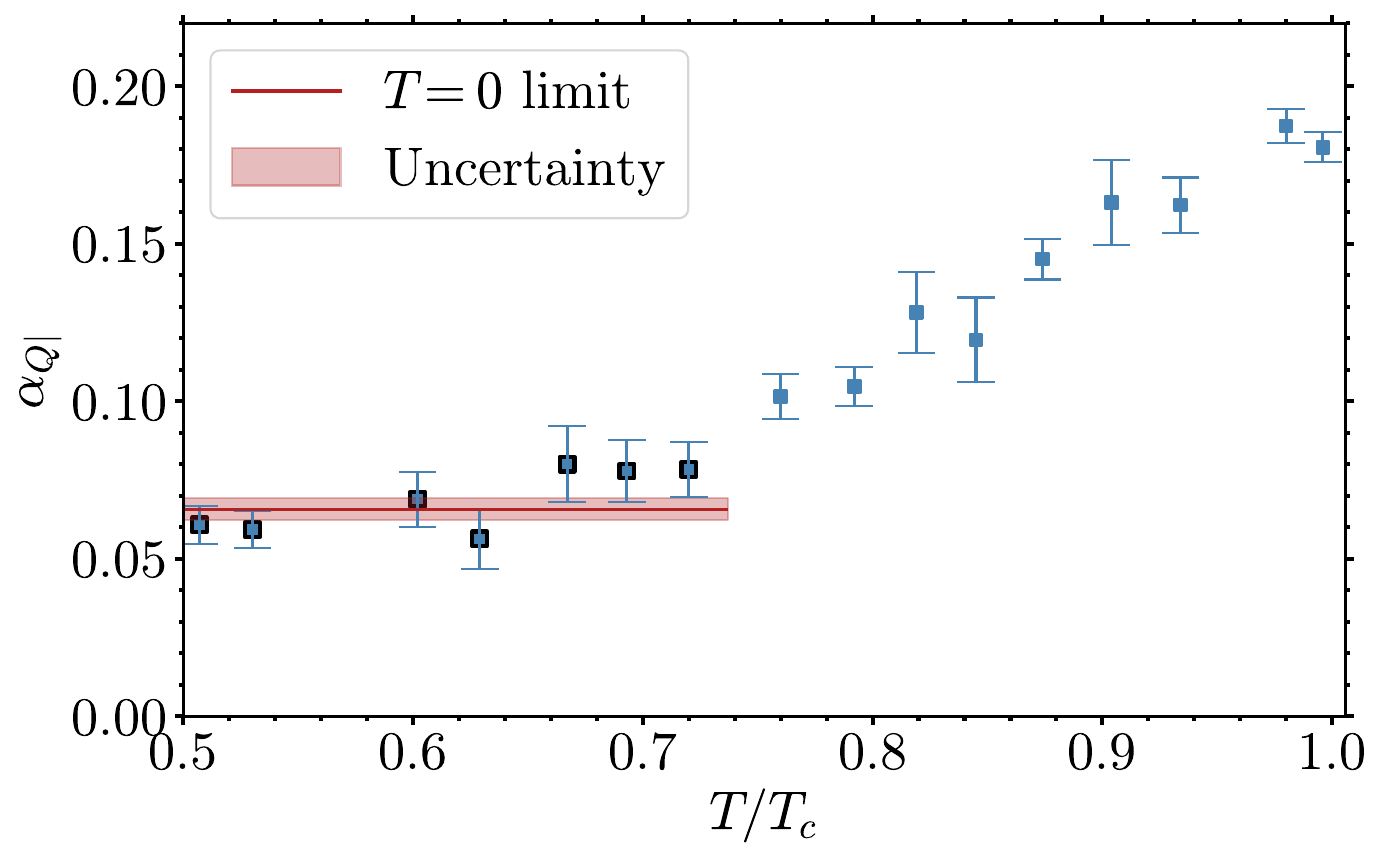}
  \caption{Coulomb coefficient $\alpha_{Q|}$ in the Cornell potential~\eqref{eq_cornell_fit} between a test quark and the mirror is shown as a function of the temperature $T$. The average of the plateau in the range of temperatures $T \leqslant 0.75 T_c$ is represented by the solid line, with the uncertainty shown by the shaded region.}
  \label{fig_alphaQ}
\end{figure}

We also examined the Coulomb coefficient $\alpha_{Q|}$ in the Cornell potential~\eqref{eq_cornell_fit} at various temperatures. The results, shown in Fig.~\ref{fig_alphaQ}, imply that the Coulomb coupling rises as a function of temperature at $T \gtrsim 0.75 T_c$. At lower temperatures, the data form a plateau, as expected. The average of the Coulomb coefficient below the temperature $T = 0.75T_c$ gives us the low-temperature limit $\alpha_{Q|} (T \to 0) = 0.066(3)$.

\section{Conclusions}

Using first-principle lattice simulations, we have found conclusive evidence of the existence of one-quark states in the confinement phase near a neutral chromoelectric mirror. A static quark is attracted to the mirror by a Cornell potential~\eqref{eq_cornell_fit}, in which a perturbative Coulomb force at short separations between a test quark and the mirror turns into a linear attractive potential at large distances. The latter fact implies that the quark is attracted to the wall by forming a confining chromoelectric string, which squeezes the chromoelectric flux originating at the quark and terminates at the wall. 

The confining quark-mirror potential implies the existence of the boundary states that we call quarkitons in the analogy with the surface excitons in condensed matter physics. The quarkitonic excitations are one-quark states near the mirror that are confined only in one dimension, normal to the surface of the mirror. The quarkitons can travel along the mirror so that in the transverse direction their motion is not confined. 

We have also found that the confining quarkitonic linear potential at large distances has a lower string tension compared to the tension of the fundamental string between a test quark and an antiquark~\eqref{eq_R}. 

Presumably, the quarkiton states can exist at the confinement-deconfinement interfaces that appear, for example, in the inhomogeneous thermodynamical ground state of rotating quark-gluon plasmas~\cite{Braguta:2023iyx, Braguta:2024zpi}.

The numerical simulations were performed on the equipment of Shared Resource Center "Far Eastern Computing Resource" IACP FEB RAS~\cite{IACPurl} and the computing cluster Vostok-1 of Far Eastern Federal University.

\begin{acknowledgments}
The work of M.N.C. was funded by the EU’s NextGenerationEU instrument through the National Recovery and Resilience Plan of Romania-Pillar III-C9-I8, managed by the Ministry of Research, Innovation and Digitization, within the project entitled ``Facets of Rotating Quark-Gluon Plasma'' (FORQ), Contract No. 760079/23.05.2023 code CF 103/15.11.2022.  The work of V.A.G. and A.V.M. was supported by Grant No. FZNS-2024-0002 of the Ministry of Science and Higher Education of Russia.
\end{acknowledgments}

\section*{Data Availability Statement}

The data that support the findings of this article are openly available \cite{data_fig}.

\bibliography{quarkiton.bib}

\end{document}